# Efficient assessment of window views in high-rise, high-density urban areas using 3D color City Information Models

Maosu Li[1*], Fan Xue[2], Anthony G.O. Yeh[3]

**Abstract**

Urban-scale quantification of window views can inform housing selection and valuation, landscape management, and urban planning. However, window views are numerous in high-rise, high-density urban areas and current automatic assessments of window views are inaccurate and time-consuming. Thus, both accurate and efficient assessment of window views is significant in improving the automation for urban-scale window view applications. The paper presents an automatic, accurate, and efficient assessment of window view indices (WVIs) of greenery, sky, waterbody, and construction using 3D color City Information Models (CIMs). The workflow includes: i) 3D semantic segmentation of photorealistic CIM and Digital Surface Model (DSM), and ii) batch computation of WVIs. Experimental results showed the estimated WVIs were more accurate (RMSE < 0.01), and the proposed method was more efficient (3.68 times faster) than Li et al.'s (2022) 2D semantic segmentation. Thus, the proposed method can facilitate large-scale WVI assessment and update in healthy high-rise, high-density urban development.

## 1 Introduction

A high-quality window view with more greenery, sky, waterbody, and fewer construction elements is treasured by urban dwellers, especially in high-rise, high-density urban areas. The recognized benefits of a high-quality window view include stress relief, life satisfaction, and productivity improvement [1, 2, 3]. In the post-Covid-19 era, the benefits of window views are further amplified for urban dwellers because of increasingly long-term indoor occupation and reduced social activities [4].

Assessment of window views can help provide quantified evidence for informed housing selection and valuation, landscape management and urban planning, and new building design. For example, renters and purchasers can holistically sort and select rooms by window view indices [5]. Property agencies can precisely value the room using quantified window view indicators [6, 7]. Urban planners and architectural designers can leverage the automatic assessment tool to compare effects of different plans and designs on window views for living environment improvement [8, 9].

However, window views are numerous especially in high-rise, high-density urban areas and often changed

[1] Department of Urban Planning and Design, The University of Hong Kong, Hong Kong SAR, China
  * Corresponding author, Email: maosulee@connect.hku.hk
[2] Department of Real Estate and Construction, The University of Hong Kong, Hong Kong SAR, China
[3] Department of Urban Planning and Design, The University of Hong Kong, Hong Kong SAR, China

in large numbers with the vertical development of neighborhoods. Fine-scale quantification of large-scale window views can consume plenty of manpower or computation resources [10, 5]. The high time cost often hinders urban planners, designers, and other decision-makers to assess and update the window view indices and examine different plans and designs for window view optimization [10, 11]. Thus, both efficient and accurate quantification of window view features is significant in advancing the window view assessment for urban-scale applications, such as housing selection and valuation, landscape management, and urban planning.

Previously, window views are often collected through onsite photography and manual modeling in psychology, built environment, and urban health fields [12, 13]. These methods are high-cost, laborious, and thus unscalable to the urban scale. Recently, simulation methods i.e., visibility analysis and view photography have been used to assess urban-scale floor-level and window views [14, 5]. Generally, traditional visibility analysis on a large-scale landscape view assessment still shows a preference for oversimplified models, such as Digital Surface Model (DSM) and polygon building models [15]. In comparison, 3D photorealistic City Information Models (CIMs) are less used due to more intensive intervisibility computations. By contrast, direct view photography on 3D photorealistic CIMs can capture photorealistic window views more easily thanks to well-developed OpenGL rendering techniques [16]. Specifically, urban-scale photorealistic window views have been generated and assessed using 3D photorealistic CIM and deep transfer learning [5]. However, the method is inaccurate for processing close-range window views and less efficient due to the repetitive segmentation of window views without reusing the shared intermediates. Thus, the next generation of assessment method needs to improve the processing efficiency for an accurate urban-scale window view assessment.

This study presents an automatic, accurate, and efficient assessment method for urban-scale window views. Specifically, we extend an assessment of the window view index (WVI) defined in Li et al.'s [5] method for color window view images generated from 3D color CIMs. A two-step workflow using photorealistic CIM and 3D semantic segmentation aims to improve both the accuracy and efficiency of the current window view assessment for urban-scale applications. The remainder is arranged as follows: research methods, experiment, discussion, and conclusion.

## 2 Research methods

Fig. 1 shows the workflow of the proposed method for efficiently assessing the urban-scale window views. The input of the method comprises a photorealistic CIM, Building Information Models (BIMs), a DSM, and a Normalized difference vegetation index (NDVI) map, as shown in Fig. 1a. Fig. 1b and 1c show the method including two steps: i) 3D semantic segmentation of photorealistic CIMs and DSM and ii) batch computation of WVIs using generated color window views. The outputs are four WVIs as shown in Fig. 1d.

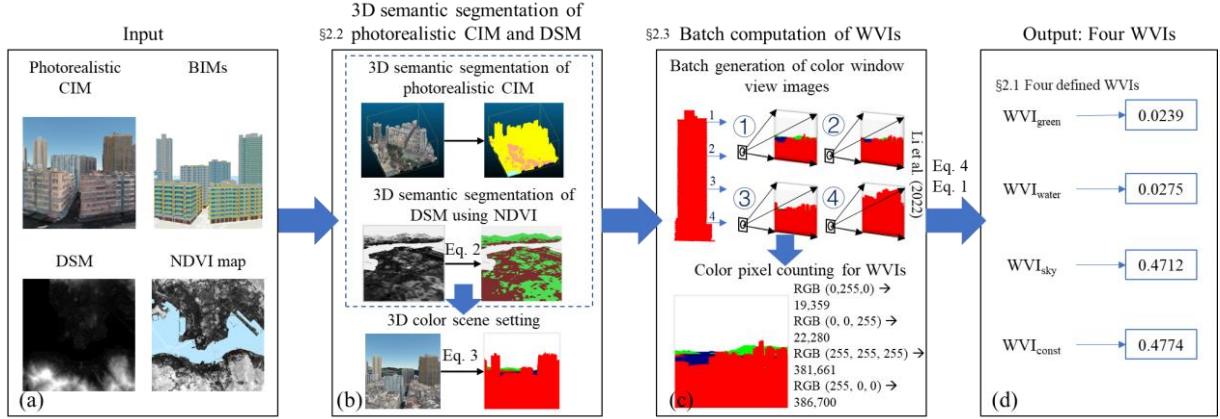

**Fig. 1.** The workflow of the proposed method. (a) Input, a proposed two-step method comprising (b) 3D semantic segmentation of photorealistic CIM and DSM and (c) batch computation of WVIs, and (d) output.

## 2.1 Definition of Window View Index (WVI)

This study applies the WVIs defined in [5] for color window view images generated from a 3D color scene as shown in Fig. 2b and 2c. Instead of using a photorealistic scene as shown in Fig. 2a, Fig. 2c and 2d show the WVI is defined as a ratio on an 8-bit RGB color view image $c$ with $n$ pixels,

$$WVI_l = |\{p|\, p \in c,\, m(p_{color}) = l\}|\,/\,n,\ l \in L,$$
$$L = \{\text{'greenery'},\ \text{'waterbody'},\ \text{'sky'},\ \text{'construction'}\}, \quad (1)$$

where $m$ is a mapping function to determine the semantic label $l$ of a pixel $p$ based on its color $p_{color}$. For example, Fig. 2c shows that the label of $p$ in green is recognized as greenery and the label of $p$ in red is recognized as construction elements. The semantic label set $L$ in this study includes greenery (green), waterbody (water), sky, and construction (const.). And $|\cdot|$ is a function to calculate the total number of pixels with the label $l$ in the image $c$. Thus, the value of $WVI_l$ ranges from 0 to 1, as shown in Fig. 2d. And the higher the value, the larger the proportion of $l$ in the window view image.

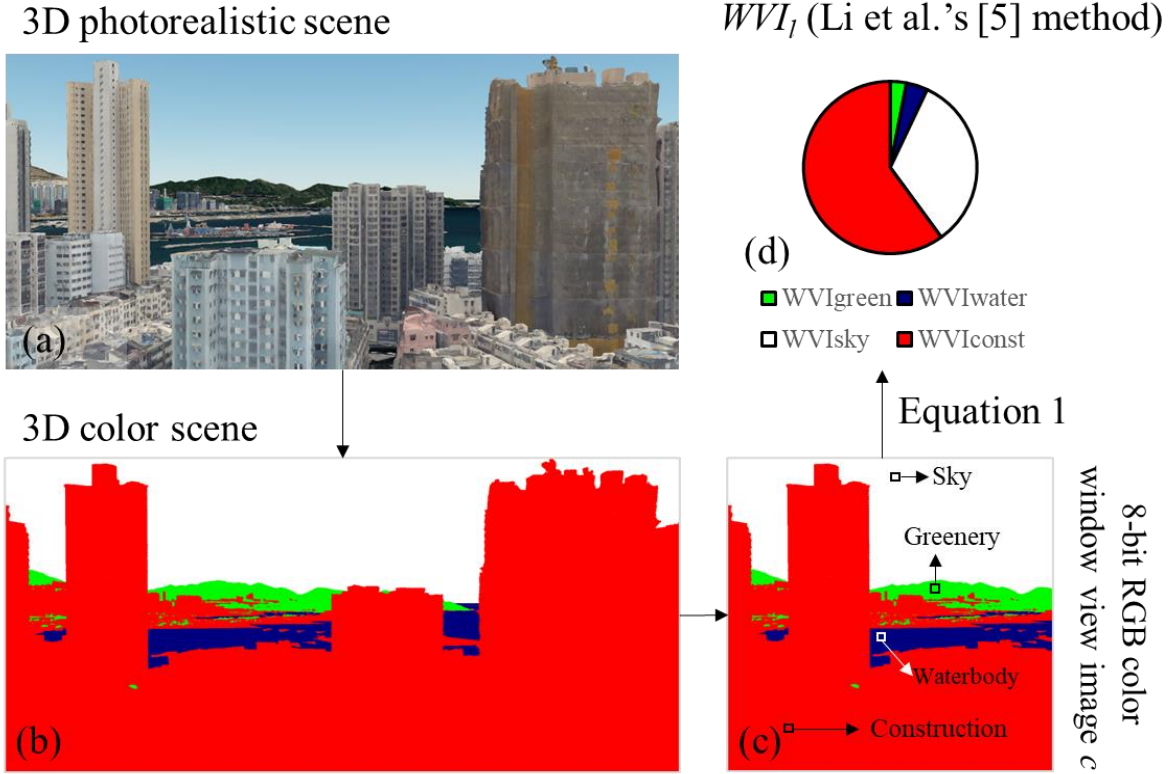

**Fig. 2.** The $WVI_l$ [5] defined on a color window view image.

### 2.2 3D segmentation of photorealistic CIM and DSM for a 3D color scene

To set up a 3D scene colored by $L$ as shown in Fig. 2b, we implement a 3D semantic segmentation on photorealistic CIM and DSM, respectively. First, a deep learning model KPConv [17] is trained to predict $l$ for every vertex of the photorealistic CIM triangles. Specifically, we evenly sample the surface of 3D photorealistic CIM into dense point clouds. Thereafter, we train the KPConv model on point clouds annotated by $L$, finetune the parameters, and then apply the well-trained model to predict the rest of the dense point clouds. Last, each point is assigned a semantic label $l$ and the triangle vertex of CIM is assigned a label same to that of the closest point.

Then, distant landscape elements represented by DSM are segmented into greenery, construction, and waterbody using the NDVI map. Since the distant view elements are rendered at a lower resolution, we use DSM to represent the distant landscape layer according to the observable saturation of measurement accuracy [10]. Specifically, the NDVI map is first segmented into categories of greenery, construction, and waterbody using Equation 2.

$$l_g = \begin{cases} \text{greenery,} & NDVI_g > 0.1, \\ \text{construction,} & 0 \leq NDVI_g \leq 0.1, \\ \text{waterbody,} & NDVI_g = \text{no data,} \end{cases} \quad (2)$$

where $g$ is the pixel of the NDVI map, and $l_g$ and $NDVI_g$ are the semantic label and NDVI value of the pixel $g$, respectively. Threshold values are set for distinguishing pixels of greenery, construction, and waterbody following guidance of local geospatial contexts. For example, we set 0.1, 0, and no data for experimental tests of Hong Kong in Section 3, following the NDVI-related report [18] of Planning Department, Hong

Kong SAR. Then, each pixel of the DSM is assigned a label *l* by segmented NDVI map through resampling and geo-registration.

Last, we color the 3D CIM, DSM, and the sky layer for a fully 3D color scene using Equation 3,

$$color(v_l) = \begin{cases} \text{RGB } (0, 255, 0), & l = \text{greenery}, \\ \text{RGB } (0, 0, 255), & l = \text{waterbody}, \\ \text{RGB } (255, 255, 255), & l = \text{sky}, \\ \text{RGB } (255, 0, 0), & l = \text{construction}, \end{cases} \quad (3)$$

where *color* is the function to map the semantic label *l* of a vertex *v* of the 3D scene into an 8-bit RGB color. For example, the vertex *v* with the semantic label *l* = greenery is colored in green, RGB (0, 255, 0), whereas the vertex with *l* = waterbody is in blue, RGB (0, 0, 255).

**2.3 Batch computation of WVIs using color view images**

This step batch computes WVIs using color window view images generated from the 3D color scene as shown in Fig. 1c. The virtual camera with a field of view at 60 degrees is first placed on each window site of the 3D color scene according to the window location information provided by the BIMs. Specifically, the virtual camera is set on the window center with *tilt* = 0, *pitch* = 0, and *heading* = (heading of the window) to capture the outside view. Then, the color view images are captured and saved in batch. Last, we set the function *m* in Equation 1 as below regarding Equation 3,

$$m(p_{color}) = \begin{cases} \text{greenery}, & p_{color} = \text{RGB } (0, 255, 0), \\ \text{waterbody}, & p_{color} = \text{RGB } (0, 0, 255), \\ \text{sky}, & p_{color} = \text{RGB } (255, 255, 255), \\ \text{construction}, & p_{color} = \text{RGB } (255, 0, 0), \end{cases} \quad (4)$$

The total number of pixels of each *l* in every window view image is summarized for computing four WVIs using Equations 4 and 1.

**3 Experiment**

**3.1 Experimental settings**

We chose 207 buildings in To Kwa Wan, Kowloon Peninsula of Hong Kong to test the feasibility of the proposed method, as shown in Fig. 3a and 3b. The study area is located in the top density zone according to the Hong Kong Planning Standards and Guidelines [19]. And the maximum building height gap is 103.15 m. There are 44,909 windows on the 207 buildings, where each building owns 217 windows on average. The 3D photorealistic CIM and DSM were collected from the Lands Department as shown in Fig. 3c and 3e [20, 21], while the window location information was extracted from BIM data shared by the Urban Renewal Authority as shown in Fig. 3d [22]. Fig. 3f shows the 30-m NDVI map collected from [23].

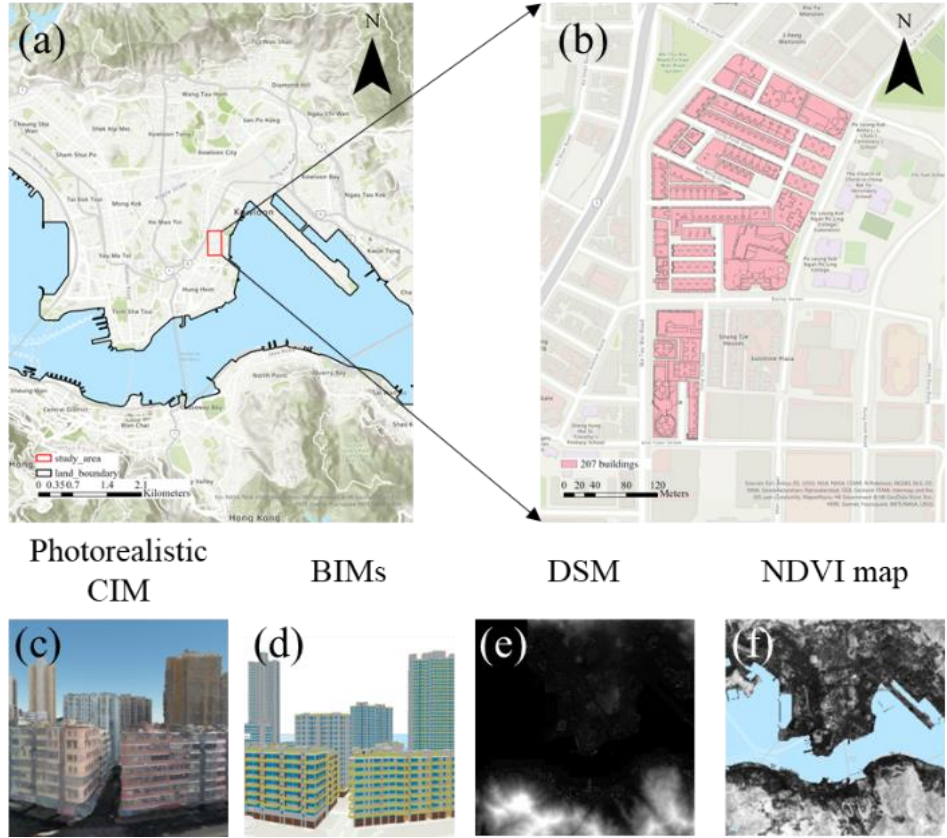

**Fig. 3.** Study area of To Kwa Wan, Hong Kong. (a) Location, (b) 207 buildings, (c) 3D photorealistic CIM, (d) BIMs, (e) DSM, and (f) NDVI map.

The computational environment was set up as follows. This study used a workstation with an Intel i9-11900K CPU (3.50 GHz, 16 cores), 64G memory, one 24G Nvidia GeForce RTX 3090 graphic card, and Windows 10 operating system (64-bit). 3D semantic segmentation was implemented in a Docker (ver. 20.10.12) container with the environment of Pytorch (ver. 1.10.0) and Python (ver. 3.7.11). We used Cesium (ver. 1.99) to set up the 3D color scene and generate the color window view images. The WVIs were quantified through a developed Python program. To compare the accuracy and efficiency, we implemented Li et al.'s [5] 2D image segmentation-based window view assessment method in the same workstation. The same version of Cesium was used to generate photorealistic window view images. And we followed the settings of [5] to quantify four WVIs using Deeplab V3+ [24] within the same Docker container. The one-off annotation of 20 tiles of sampled point cloud for our 3D semantic segmentation and 110 window view images for Li et al.'s 2D semantic segmentation consumed about 10 person-hours each. In addition, we manually annotated another 100 window view images selected in Section 3.2 to test the accuracy of both assessment results. Distant landscape elements beyond 2,000 m were colored using DSM and NDVI referring to [10]. Same to Li et al.'s window view settings, window view images with 900 × 900 pixels were generated in batch.

### 3.2 Experimental results

Overall, the performance of the trained KPConv in our method reached a mIoU of 0.91 for the segmentation of greenery, waterbody, and construction from the 3D photorealistic CIM. The values of $R^2$ for estimating four WVIs using Li et al.'s 2D method were above 0.95. Specifically, we randomly generated 100 window view images from the 207 buildings for the accuracy and efficiency test. Table 1 shows that the assessment results of our method reached a higher accuracy (RMSE < 0.01) than those of Li et al.'s 2D method. Observable reason was more accurate segmentation of close-range window views using our method as shown in Figure 4. Figure 4b shows there existed misclassification of close-range construction elements into other landscape elements such as vegetation using Li et al.'s 2D method. By contrast, the results of our 3D method were more satisfactory due to the holistic 3D segmentation of the photorealistic CIM and DSM. Fig. 4c shows close-range view elements were correctly recognized as construction.

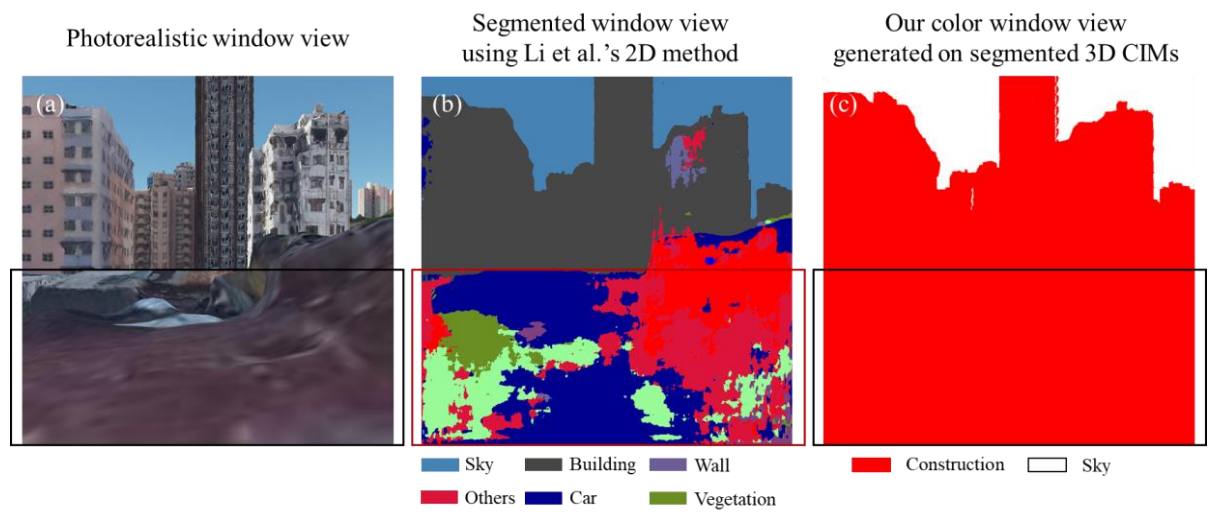

**Fig. 4.** Comparison of the segmentation accuracy against close-range view elements using two methods. (a) Example photorealistic window view generated from CIMs, (b) segmented window view using Li et al.'s [5] 2D method, and (c) our color window view generated on segmented 3D CIMs.

**Table 1.** Comparison of assessment accuracy of two methods on 100 test windows

|  | RMSE | | |
| --- | --- | --- | --- |
|  | Li et al.'s 2D method | Our 3D method | Improvement |
| $WVI_{green}$ | 0.0283 | 0.0059 | 79.15% |
| $WVI_{water}$ | 0.0243 | 0.0048 | 80.25% |
| $WVI_{sky}$ | 0.0098 | 0.0044 | 55.10% |
| $WVI_{const.}$ | 0.0405 | 0.0092 | 77.28% |
| Average | 0.0257 | 0.0061 | 76.26% |

Meanwhile, Table 2 shows the processing of 100 window view images using our method saved 73% of the total time cost compared to Li et al.'s 2D method. The window view generation of our method was efficient (3.59 times faster) without the preparation and rendering of the 3D CIM texture images. The quantification of WVIs improved 81% of the efficiency by avoiding the repetitive 2D segmentation of window view images. In summary, the results in Tables 1 and 2 confirmed that our method was both accurate and efficient for urban-scale window view assessment.

Table 2. Comparison of computational time of the two methods (average of 100 windows)

| Step | Processing | Li et al.'s 2D method | Our 3D method | Improvement |
|---|---|---|---|---|
| 1 | Window view generation | 1.94 s | 0.54 s | 72% |
| 2 | Quantification of WVIs | 0.16 s | 0.03 s | 81% |
| | Total | 2.10 s | 0.57 s | 73% |

## 4 Discussion

Both accurate and efficient window view assessment method is important for examining and updating urban-scale window view indices, which can benefit precise housing valuation and selection, landscape management and urban planning, and new building design. For example, an efficient assessment method can encourage urban planners and property agencies to quantify and timely update the latest window view indices of greenery, waterbody, sky, and construction at the urban scale. The updated accurate window view indices of greenery, waterbody, sky, and construction can help urban planners to identify the urban areas with less nature exposure for prioritized planning and design practices such as planning more blue-green spaces [11]. Property agencies can use large-scale accurate window view indicators for precise valuation. Instead of the previous qualitative judgment, housing purchasers and renters can select rooms with a satisfactory view by comparing and sorting quantified view indicators of rooms holistically. In addition, leveraging an accurate and efficient assessment tool, architectural designers can compare the effects of different plans and design drafts on window views with a low time cost.

Recently, the urban-scale window view assessment has been implemented at the urban scale using 3D photorealistic CIM and deep transfer learning. However, the method is still inaccurate and inefficient for processing urban-scale window views. The root reasons are inaccurate 2D segmentation of close-range window view elements, the high workload of texture rendering, and 2D segmentation with repeated computations. For example, the close-range window view elements can be segmented into incorrect categories due to the limited resolution of the CIM texture images. The generation of window view images is less efficient due to the high workload of CIM texture rendering. Landscape elements shared by numerous windows are repetitively segmented for every window view without reusing the intermediates.

The contribution of this paper is to present a novel method to improve the accuracy and efficiency of window view assessment. A 3D scene colored by four view types (e.g., greenery, waterbody, sky, and construction) using 3D CIMs and 3D semantic segmentation is set up to avoid the close-range view segmentation and repetitive computation, which effectively improves the accuracy and efficiency for an urban-scale window view assessment. Experimental results show that the assessment results of our method were more accurate (RMSE < 0.01) and the proposed method was 3.68 times faster than Li et al.'s [5] 2D semantic segmentation.

There still exist some limitations in the study. First, a full 3D color scene needs to be prepared for assessing four WVIs. Small-scale quantification, e.g., for one or two window views may not afford the large-scale but one-off preprocessing cost. In other words, the proposed method is born for a large-scale window view assessment. In addition, since neighborhood windows especially on the upper stories of high-rise buildings often share similar views, the repetitive quantification regardless of the spatial context may hinder the efficiency of the urban-scale assessment. Thus, future directions include simplification of preprocessing of 3D CIMs and mining of window view patterns of different types, locations, and storeys of buildings and blocks for a more efficient urban-scale window view assessment.

## 5 Conclusion

Window views are numerous and changed in large numbers with the vertical development of the urban landscapes especially in high-rise, high-density urban areas. The current methods are inaccurate and inefficient to process window views at the urban scale due to inaccurate segmentation and high workloads of repetitive computation. Thus, automatic, accurate, and efficient window view assessment is significant for assessing and updating thousands of window view indices for urban-scale applications such as housing selection and valuation, landscape management, and urban planning.

This study extends the assessment of four defined Window View Indices (WVIs) of greenery, waterbody, sky, and construction for 3D color window view images generated from a 3D color scene. The 3D scene set up by photorealistic CIM and DSM is first colored by the four types of landscape elements using 3D semantic segmentation. Then, 3D color window view images are generated from the 3D color scene by placing the virtual camera on the window sites. Last, we programmatically summarize the pixels in different colors for computing four WVIs. Experimental results in To Kwa Wan, Hong Kong showed that our assessment results were more accurate (RMSE < 0.01) and the proposed method was 3.68 times faster than the state-of-the-art method. The both accurate and efficient assessment of window views can be useful for encouraging property agencies, urban planners, and other decision-makers to timely assess and update urban-scale window view indices for housing valuation and selection, landscape urban management, and urban planning. Future work includes simplifying the preprocessing of 3D CIMs and pattern mining of window views for a more efficient urban-scale window view assessment.